\documentclass[pra,floatfix,showpacs,amsmath,twocolumn]{revtex4-1}

\makeatletter
\let\@ORGREVTEXendnotemark\@endnotemark
\let\@ORGREVTEX@makefnmark@cite\@makefnmark@cite
\def\@endnotemark{\bgroup\@fileswfalse\@ORGREVTEXendnotemark\egroup}
\def\@makefnmark@cite{\bgroup\@fileswfalse\@ORGREVTEX@makefnmark@cite\egroup}
\makeatother

\usepackage{amssymb}
\usepackage[mathscr]{eucal}
\usepackage{graphicx}

\usepackage{braket}
\usepackage{enumerate}
\usepackage{subfigure}

\usepackage{mciteplus}
\mciteErrorOnUnknownfalse

\usepackage{color}

\def\tr{\mbox{Tr}}

\def\vecmap{\mbox{vec}}

\newcommand{\1}{1\hspace{-0.243em}\text{l}}
\newcommand{\bvec}[1]{\mathbf{#1}}

\def\A{{\bvec{A}}}
\def\B{{\bvec{B}}}

\begin{document}

\title{A matrix product state based algorithm for determining dispersion
relations of quantum spin chains with periodic boundary conditions}

\author{B. Pirvu$^1$, J. Haegeman$^2$, F. Verstraete$^1$}
\affiliation{$^1$Fakult\"at f\"ur Physik, Universit\"at Wien,
Boltzmanngasse 5, A-1090 Wien, Austria\\
$^2$Ghent University, Department of Physics and Astronomy, Krijgslaan 281-S9, B-9000 Ghent, Belgium}

\date{\today}

\begin{abstract}
We study a matrix product state (MPS) algorithm to approximate excited
states of translationally invariant quantum spin systems with
periodic boundary conditions. By means of a momentum eigenstate
ansatz generalizing the one of \"Ostlund and Rommer
\cite{rommer-ostlund-1995,*rommer-ostlund-1997}, we separate
the Hilbert space of the system into subspaces with different
momentum. This gives rise to a direct sum of effective Hamiltonians,
each one corresponding to a different momentum, and we determine
their spectrum by solving a generalized eigenvalue equation. Surprisingly,
many  branches of the dispersion relation are approximated to a very
good precision. We benchmark the accuracy of the algorithm by
comparison with the exact solutions of the quantum Ising and the
antiferromagnetic Heisenberg spin-$1/2$ model.

\end{abstract}

\maketitle



\section{Introduction}

Recently we have presented an algorithm~\cite{me-2010-PBCI} for the
approximation of the ground state of translationally invariant (TI) spin
chains with periodic boundary conditions (PBC) my means of TI MPS.
In this work we will use the ground states
obtained in~\cite{me-2010-PBCI} as the basis of an ansatz for
excited states with definite momentum.
We will consider only spin chain Hamiltonians that are translationally
invariant thereby fulfilling $[H,T]=0$ where $T$ is the translation
operator that shifts the lattice by one site.
Furthermore, as we will deal with finite chains in the following,
it means that there is no well defined momentum operator for our systems.
Nevertheless we can classify translationally invariant states
by their quasi-momentum which is defined in terms of the their
eigenvalue with respect to $T$. This definition is sensible
since in the thermodynamic limit, if we keep the chain length fixed,
the lattice spacing becomes infinitesimally small
and the quasi-momentum becomes identical to the momentum, which is
then well defined. For convenience, we will use
the term \emph{momentum} when we actually refer to the quasi-momentum.
This should not cause any confusion since we will only deal with
quasi-momenta throughout this work.

Since $H$ and $T$ commute, they can be diagonalized simultaneously.
This suggests that any variational ansatz based on eigenstates of
the translation operator will be well suited
to define families of states within which minimization with
respect to some variational parameters will yield \emph{momentum}
eigenstates with minimal energy.
Formulating this observation in terms of an MPS-based ansatz has
led in the past to some very interesting results about excitation spectra.
The first approach in this direction has been made in
\cite{rommer-ostlund-1995}
where the main result
is the celebrated insight that the fixed point of the density matrix
renormalization group (DMRG) can be written as an MPS.
In addition to this, based on
the MPS that is obtained for the ground state of the infinite Heisenberg
spin-$1$ chain, the authors suggest a variational ansatz for excitations
with definite \emph{momentum}.
Since the translationally invariant MPS they start
with is an approximation of the ground state in the thermodynamic limit,
their ansatz for excitations is only well suited in the limit $N\to\infty$.
For finite chains, the idea of using \emph{momentum} eigenstates for
the diagonalization of TI Hamlitonians has been used in~\cite{porras-2006}
in order to obtain a few of the lowest branches of excitations
for the bilinear-biquadratic (BB) spin-1 chain. The resulting state is a
TI superposition of a special class of tensor network states, which can
be viewed as an extension of MPS with PBC~\cite{frank-2004a} to
states that can accommodate multipartite entanglement.
Even though the multipartite entanglement is a nice feature which yields
a better variational ansatz in the cases when the approximated states have
that special entanglement structure (in~\cite{porras-2006} one has in
addition to the usual maximally entangled virtual bonds between nearest
neighbors a virtual GHZ state connecting all sites) we will not adopt it
in our present ansatz. Furthermore we would like to point out that the
individual MPS tensors produced by the minimization procedure
in~\cite{porras-2006} are not TI, only their superposition is.

Recent results~\cite{me-2010-PBCI} on the
approximation of ground states of TI PBC Hamiltonians opened up the
possibility of unifying the ideas
from~\cite{rommer-ostlund-1995} and~\cite{porras-2006} in order to obtain
an algorithm for excitations with definite \emph{momentum} in which only one
local tensor has to be determined, thereby avoiding the usual sweeping
procedure and the associated factor $N$ in the computational cost.
One of the main features of TI MPS is the fact that the
tensor network that has to be contracted for the computation of expectation
values contains big powers of a so-called transfer
matrix~\cite{fannes-1992}.
For non-critical systems the eigenvalues of this transfer matrix
usually decay rapidly enough s.t. big powers thereof can be
accurately approximated by considering only a few dominant eigenvectors.
In these cases the computational cost for the evaluation of expectation
values for systems with PBC
can be reduced significantly from $O(D^5)$ to $O(D^3)$, where $D$
denotes the virtual bond dimension of the MPS. For critical systems however
the eigenvalues of the transfer matrix decay much slower and the
algorithm that must be employed in order to obtain the optimal approximation
within the class of MPS with fixed $D$ has a scaling that depends in a
not yet fully understood way~\cite{me-2010-PBCI} on $D$, $N$ and on the
universality class of the simulated model.

The ansatz we present in this work is based on TI-MPS and thereby
all computed quantities will contain big powers of the transfer matrix.
We would like to emphasize that the computational cost can be reduced
by a factor of $D^2$ only in the case of non-critical systems. For critical
systems the full contraction of tensor networks (i.e. without using any
approximations of the transfer matrix) will turn out to have a more
favorable overall scaling of the computational cost.
Details on why this is the case and on the scaling of the computational cost
can be found in the next section.

\section{Overview}
\label{sec:overview}

Due to $T^N=\1$, the translation operator $T$ that shifts a state on a
PBC lattice with $N$ sites by $1$ site is the generator of the cyclic group
of order $N$. Hence its eigenvalues $\tau_k$ must be
roots of the unity i.e. $\tau_k=e^{-i k \frac{2\pi}{N}}$ with
integer $k\in [0,N-1]$.
An ansatz for eigenstates of $T$ with eigenvalue $e^{-i k \frac{2\pi}{N}}$
is obviously given by

\begin{equation}\label{eq:ansatz}
  \ket{\psi_k(\B)}=\sum_{n=0}^{N-1} \frac{1}{\sqrt{N}}
  e^{i\frac{2\pi k n}{N}} T^n \ket{\phi_\A(\B)}
  \,\,\,.
\end{equation}
\noindent
Henceforth we will refer to states of the form (\ref{eq:ansatz}) as
Bloch states.
Note that we have used the convention that $T$ is the operator that realizes
a translation by one site to the right s.t.
$T\ket{\phi(i_1,i_2,\dots,i_N)}=\ket{\phi(i_N,i_1,i_2,\dots,i_{N-1})}$.
The state $\ket{\phi_\A(\B)}$ can in principle be any arbitrary state,
but in order to exploit the advantages of TI MPS, we choose

\begin{equation}\label{eq:ansatz_phi}
  \ket{\phi_\A(\B)}=
  \sum_{i_1,\dots i_N=1}^{d}
  \tr\big(B_{i_1} A_{i_2} \dots A_{i_N}\big)
  \ket{i_1 i_2 \dots i_N}
\end{equation}
\noindent
with identical matrices $A_i$ on all sites except the first one.
We will choose the $A_i$ to be the matrices corresponding
to the best TI MPS ground state approximation for a given model.
We emphasize that the $A_i$ remain fixed throughout the entire simulation.
This is the reason why we have omitted them from our labeling convention
for the Bloch states $\ket{\psi_k(\B)}$.
We have used bold letters in order to denote objects that are
obtained if one rearranges the components of three indexed
MPS tensors into vectors, i.e.
$\bvec{A}:=\vecmap( A_{i\phantom{\alpha}\beta}^{\phantom{i}\alpha} )$.
After fixing the \emph{momentum} $k$, the Bloch states $\ket{\psi_k(\B)}$
will depend only on the tensors $\B$ which will define the variational
manifold.


Our ansatz for Bloch eigenstates differs slightly from the ones
presented in Refs.~\cite{rommer-ostlund-1995,porras-2006,chung-2009}
although it is conceptually very similar. An important feature
of all these approaches is the reduction of the dimension of the
problem by a factor $N$. This is reached by effectively
projecting the original problem into the subspace with fixed
\emph{momentum} $k$ and minimizing the energy within the variational
manifold spanned by the free parameters in the ansatz. In our
case these free parameters are the components $\B$ of an MPS tensor.
As it is always the case with MPS algorithms, one must eliminate
the ambiguities arising from the MPS representation by fixing the gauge.
Here this is done by starting with certain tensors $\A$ in
(\ref{eq:ansatz_phi}) and not changing them throughout the entire
minimization procedure.
This automatically fixes the gauge of the tensors $\B$ as they are
surrounded on both sides by $\A$.

\section{The algorithm}
\label{sec:algorithm}

Ansatz (\ref{eq:ansatz}) defines a class of variational states
for the lowest energy states with fixed \emph{momentum}.
The energy is a quadratic expression in the tensor $\B$ and thereby,
as it is usually the case in MPS based algorithms, minimizing

\begin{equation}\label{eq:variational_problem}
  E_0(k)
        =\min_{\bvec{B}\in\mathcal{C}^{dD^2}}
         \frac{\bra{\psi_k(\bvec{B})} H \ket{\psi_k(\bvec{B})}}
              {\braket{\psi_k(\bvec{B})|\psi_k(\bvec{B})}}
        \,\,\,,
\end{equation}
\noindent
is equivalent to solving a generalized eigenvalue equation

\begin{equation}\label{eq:generalized_eigenvalue_problem}
  H_{eff}(k) \bvec{B}_i(k) = E_i(k) N_{eff}(k) \bvec{B}_i(k)
\end{equation}
\noindent
where $H_{eff}(k)$ is defined by

\begin{equation}\label{eq:Heff_def}
  \B^{\dagger} H_{eff}(k) \B
  := \bra{\psi_k(\bvec{B})} H \ket{\psi_k(\bvec{B})}
\end{equation}
\noindent
and $N_{eff}(k)$ by

\begin{equation}\label{eq:Neff_def}
  \B^{\dagger} N_{eff}(k) \B
  := \braket{\psi_k(\bvec{B})|\psi_k(\bvec{B})}
  \,\,\,.
\end{equation}
\noindent

The eigenvector corresponding to the smallest eigenvalue
$E_0(k)$ yields then the tensor $\B_0(k)$ that when plugged into our
ansatz~(\ref{eq:ansatz}) gives the \emph{momentum}-$k$ state with
minimal energy.
Note that the variational principle guarantees that only the Bloch state
(\ref{eq:ansatz}) with lowest energy is the best approximation to
the exact eigenstate with that \emph{momentum}
within the subspace spanned by our ansatz states.
However if the lowest energy state is approximated accurately, due to the fact
that the other $\B_i(k)$ are orthogonal to $\B_0(k)$, the next solution
$\B_1(k)$ has a good chance to be close to the next higher energy state
with that momentum.
In fact it will turn out that quite a few
of the higher energy solutions of
(\ref{eq:generalized_eigenvalue_problem}) are
good approximations to low energy states with fixed \emph{momentum}.
Their precision is most of the time surprisingly good given the fact that
the variational principle does not hold for these states. The quality
of these solutions depends strongly on the bond dimension $D$, the
chain length $N$ and the model under consideration.

\begin{figure}[h]
  \begin{center}
    \includegraphics[width=1.0\columnwidth]{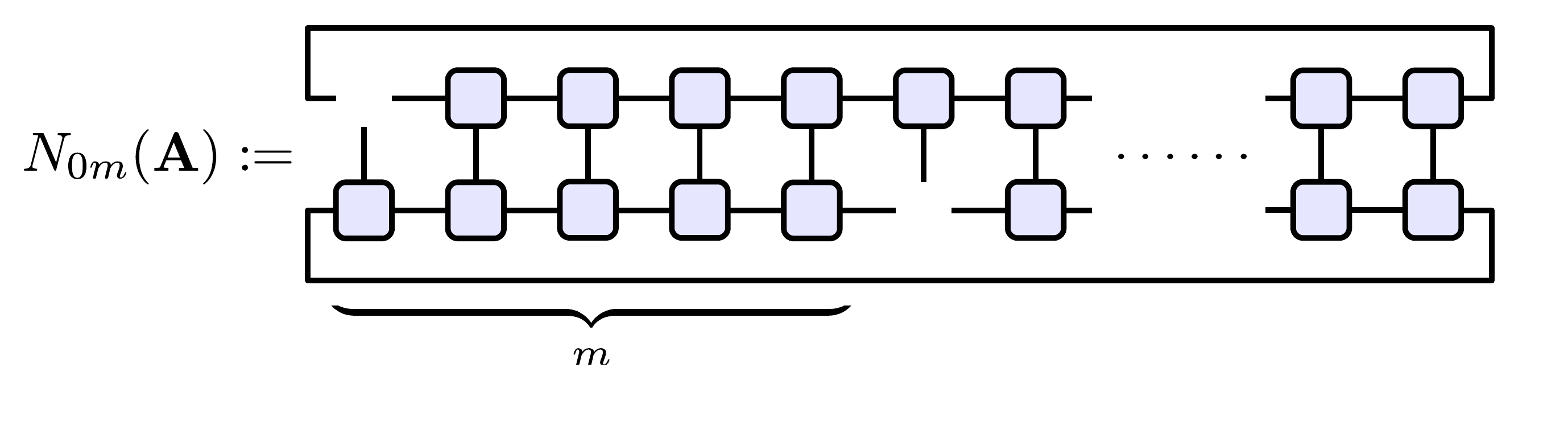}
  \end{center}
  \caption{
    (Color online).
    Definition of $N_{0m}(\A)$ as the norm of a TI MPS determined
    by the tensor $\A$.
  }
\label{fig:N_0m}
\end{figure}

The bottleneck of our method is the computation of the effective
matrices $H_{eff}(k)$ and $N_{eff}(k)$. Let us first
consider $N_{eff}(k)$ since it is the slightly simpler one.
It reads

\begin{equation}\label{eq:Neff}
\begin{split}
  \B^{\dagger} &N_{eff}(k) \B =\\
  &=\frac{1}{N} \sum_{m,n=0}^{N-1} e^{i\frac{2\pi k (n-m)}{N}}
  \bra{\phi_\A(\B)} T^{(n-m)} \ket{\phi_\A(\B)}\\
  &=\sum_{\bar{n}=0}^{N-1} e^{-i\frac{2\pi k \bar{n}}{N}}
  \bra{\phi_\A(\B)} T^{-\bar{n}} \ket{\phi_\A(\B)}\\
  &= \B^{\dagger} \bigg[ \sum_{m=0}^{N-1} e^{-i\frac{2\pi k m}{N}}
  \cdot N_{0m}(\A) \bigg] \B
\end{split}
\end{equation}
\noindent
where $N_{0m}(\A)$ is a tensor network
resembling the norm of a TI MPS with empty slots $0$ and $m$
(see figure~\ref{fig:N_0m}).
To get from the second to the third line we have used the
fact that due to the PBC only the relative distance between $n$ and
$m$ plays a role. In the last line we have merely renamed the
summation index and introduced the quantity $N_{0m}(\A)$.
Thus in order to obtain $N_{eff}(k)$ we have to compute the contraction of
the $N$ tensor networks $N_{0m}(\A)$ and then take
the sum of these terms after weighting each one of them with the
corresponding phase factor. The computational cost for the contraction
of each tensor network is $O(D^6)$ s.t. the overall cost for computing
$N_{eff}(k)$ is $O(ND^6)$.

\begin{figure}[h]
  \begin{center}
    \includegraphics[width=1.0\columnwidth]{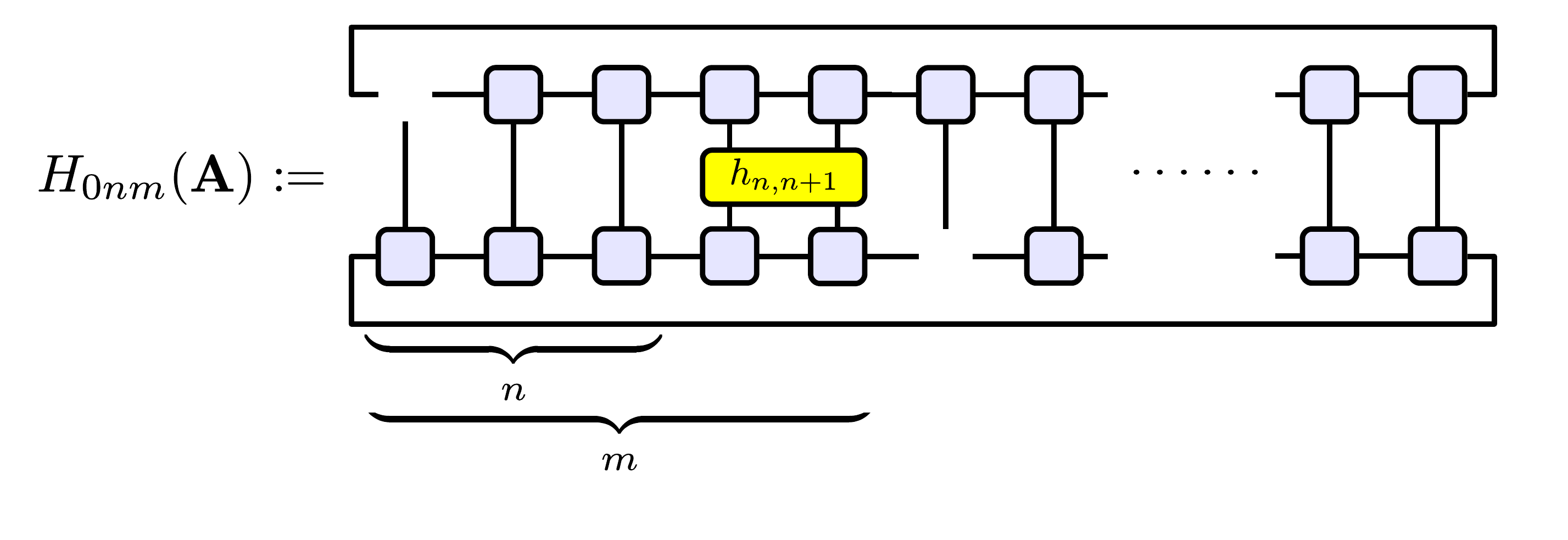}
  \end{center}
  \caption{
    (Color online).
    Definition of $H_{0nm}(\A)$ as the expectation value of a
    two-site operator with respect to a TI MPS determined
    by the tensor $\A$.
  }
\label{fig:H_0nm}
\end{figure}

$H_{eff}(k)$ is constructed very much in the same spirit.
First, due to the translational invariance of the Hamiltonian we can write

\begin{equation}\label{eq:H_TI}
  H=\sum_{l=0}^{N-1} h_{l,l+1} = \sum_{l=0}^{N-1} T^{l} h_{01} T^{-l}
\end{equation}
\noindent
where $h_{01}$ is the term acting between the first two sites of the
chain. Note that in~(\ref{eq:H_TI}) we have restricted ourselves
to nearest neighbor Hamiltonians since this is the type of models
we will treat numerically in this work. Generalizing the ideas developed
here to any local Hamiltonian is straightforward.
With~(\ref{eq:H_TI}), $H_{eff}(k)$ reads

\begin{equation}\label{eq:Heff_01}
\begin{split}
  &\B^{\dagger} H_{eff}(k) \B =\\
  &=\frac{1}{N} \sum_{l,m,n=0}^{N-1} e^{i\frac{2\pi k (n-m)}{N}}
    \bra{\phi_\A(\B)} T^{l-m} h_{01} T^{n-l} \ket{\phi_\A(\B)}\\
  &=\frac{1}{N} \sum_{l=0}^{N-1} \sum_{\bar{m},\bar{n}=-l}^{N-1-l}
     e^{i\frac{2\pi k (\bar{n}-\bar{m})}{N}}
     \bra{\phi_\A(\B)} T^{-\bar{m}} h_{01} T^{\bar{n}} \ket{\phi_\A(\B)}
  \,\,\,.
\end{split}
\end{equation}
\noindent
Again, due to the fact that the $\bar{m}$ and $\bar{n}$ sums run over
all $N$ sites of a PBC chain, it is irrelevant where they begin s.t.
the $l$ sum merely yields a factor $N$. We rename the summation
indices for convenience and obtain

\begin{equation}\label{eq:Heff_02}
\begin{split}
  &\B^{\dagger} H_{eff}(k) \B =\\
  &=\sum_{m,n=0}^{N-1} e^{i\frac{2\pi k (n-m)}{N}}
    \bra{\phi_\A(\B)} T^{n-m} T^{-n} h_{01} T^{n} \ket{\phi_\A(\B)}\\
  &=\sum_{n=0}^{N-1} \sum_{\bar{m}=n}^{n-N+1}
    e^{-i\frac{2\pi k \bar{m}}{N}}
    \bra{\phi_\A(\B)} T^{-\bar{m}} h_{n,n+1} \ket{\phi_\A(\B)}\\
  &=\B^{\dagger} \bigg[ \sum_{m,n=0}^{N-1}
    e^{-i\frac{2\pi k m}{N}}
    \cdot H_{0nm}(\A) \bigg] \B
\end{split}
\end{equation}
\noindent
where $H_{0nm}(\A)$ is a tensor network
resembling the expectation value of an operator acting on the sites
$n$ and $n+1$ with respect to a TI MPS where the slots $0$ and $m$
have been left open (see figure~\ref{fig:H_0nm}).
The computational cost for the contraction
of each tensor network is again $O(D^6)$ but now we have a total
of $N^2$ summands s.t. the overall cost for computing
$H_{eff}(k)$ is $O(N^2D^6)$. Note that to obtain $H_{eff}(k)$ is
computationally the most expensive part of our algorithm so we
can say that the overall computational cost scales like $O(N^2D^6)$.

\begin{figure*}[ht]
  \begin{center}
    \includegraphics[width=1.0\textwidth]{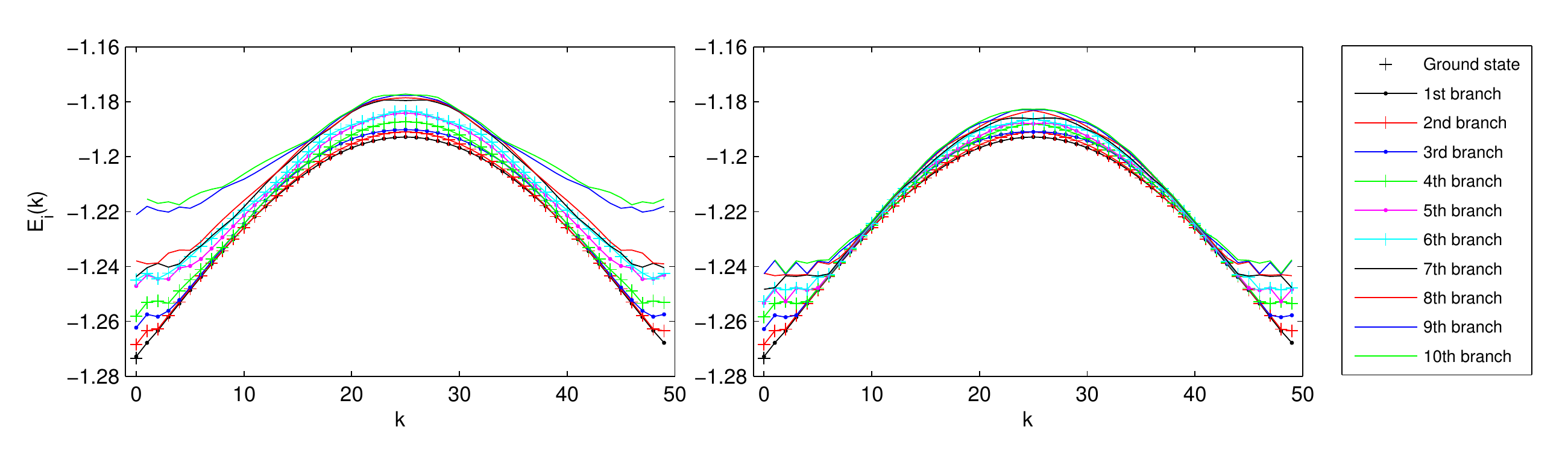}
  \end{center}
  \caption{
    (Color online).
    Lowest ten branches of the excitation spectrum for a critical
    Ising chain with $N=50$. Left: $D=8$.
    Right: $D=32$.
  }
\label{fig:IS_Ei_g10_N50}
\end{figure*}

\subsection{Overall scaling of the computational cost}\label{sec:comp_cost}

At first sight the cost seems horrible for a 1D-algorithm.
Let us however have a closer look at what we get for this price.
First of all note that if we compute the sets of matrices
$\{N_{0m}(\A)\}$ and $\{H_{0nm}(\A)\}$ for $n,m\in[0,N-1]$ and store
these, we can obtain the $H_{eff}(k)$ and $N_{eff}(k)$ for
all $k$ trivially by just building the appropriately
weighted sums.
For each of these $k$ we then have to solve the generalized
eigenvalue equation~(\ref{eq:generalized_eigenvalue_problem}).
Since $H_{eff}(k)$ and $N_{eff}(k)$ are small $dD^2\times dD^2$ matrices
solving~(\ref{eq:generalized_eigenvalue_problem}) does not represent
any difficulty and can be done using any standard library eigenvalue solver.
Each eigenvalue problem leads to $dD^2$ orthonormal vectors $\B_i(k)$
which plugged into the ansatz~(\ref{eq:ansatz}) yield
$dD^2$ states. Thus computing the sets $\{N_{0m}(\A)\}$ and
$\{H_{0nm}(\A)\}$ only once supplies immediately us with $NdD^2$ states!
By comparing our numerical results to exactly solvable
models we will show that the low energy states obtained
in this way are very accurate. This means that in terms of computational
time per state our algorithm performs quite well.

The computational bottleneck at the moment is that
we have to store $N^2$ $dD^2\times dD^2$ matrices in the memory.
With the present implementation, for a chain with $N=100$ sites,
we can go up to $D=32$. For larger $N$ simulations we have to settle
for smaller $D$. It is however straightforward how this boundary can
be pushed considerably towards larger $D$. First, instead of keeping
all matrices in the memory, one can write them to the hard disk after
computing each of them. Second, since the $\{N_{0m}(\A)\}$ and
$\{H_{0nm}(\A)\}$ are independent, one can parallelize their computation.

Thus the conceptual bottleneck becomes the contraction of the tensor
networks $\{N_{0m}(\A)\}$ and $\{H_{0nm}(\A)\}$. For non-critical
systems big powers of the transfer matrix can be well approximated by
a few of its dominant eigenvectors~\cite{me-2010-PBCI} and the contraction
of \emph{most} of the $\{N_{0m}(\A)\}$ can be done with the computational
cost $O(n^2 D^3)$ while that of \emph{most} of the $\{H_{0nm}(\A)\}$ with
the cost $O(n^3 D^3)$. Here $n$ represents the dimension of the subspace
within which we approximate the powers of the transfer
matrix~\cite{me-2010-PBCI}. This cannot be done however for critical
systems where in principle $n$ may grow as big as $D^2$ thereby
yielding a much worse scaling than the naive $O(D^6)$.
Note that since $\{N_{0m}(\A)\}$ and $\{H_{0nm}(\A)\}$ are
\emph{open} tensor networks the $O(D^5)$ contraction
scheme~\cite{frank-2004a} that works
for expectation values (i.e. \emph{closed} tensor networks) cannot be
applied here. Additionally, even if we restrict ourselves
to non-critical systems, not \emph{all} of the $\{N_{0m}(\A)\}$
and $\{H_{0nm}(\A)\}$ can be computed with the cost that scales
like $D^3$: if the distance between the \emph{open} slots is not
big enough, we cannot use the approximation for big powers of
the transfer matrix between the slots, and we are back to exact
contraction for this portion of the chain which in the case of
$\{N_{0m}(\A)\}$ leads to the overall scaling $O(nD^5)$ and
in the case of $\{H_{0nm}(\A)\}$ to the scaling $O(n^2D^5)$.
Thus the very naive exact contraction procedure that we use is
not so bad after all in this case even if it scales like $O(D^6)$.

There is one more subtlety we would like to point out here.
It turns out that the matrix $N_{eff}(k)$ is always singular which
presents a problem when we try to solve the generalized eigenvalue
equation~(\ref{eq:generalized_eigenvalue_problem}) since the solution
involves the inverse $N_{eff}^{-1}(k)$. We can circumvent this
problem by solving~(\ref{eq:generalized_eigenvalue_problem})
within the nonsingular subspace like it has been done
in~\cite{rommer-ostlund-1995}. Eigenvectors associated to the zero
eigenspace of $N_{eff}(k)$ will result in physical states
$\ket{\psi_k(\B)}=0$, i.e. these are states of zero length in the Hilbert
space. Any physical operator will produce a zero when acting on these
states. In particular, the effective Hamiltonian $H_{eff}(k)$ will also
have zero eigenvalues for the same eigenvectors, and we do not loose any
information by restricting to the nonsingular subspace. The dimension of the
zero eigenspace can be shown to be $D^2(d-1)$ for $k\neq 0$ and
$D^2(d-1)+1$ for $k=0$ as we demonstrate in \cite{juthoEXC-2011}.
The tricky point is that for some models the strictly non-zero
eigenvalues of $N_{eff}(k)$ become so small that they yield the
generalized eigenvalue problem ill conditioned.
In general this behavior does not occur for small $D$.
For big $D$ or in certain regions of the phase diagram however
the nonsingular eigenvalues become so small that it is
hard to distinguish numerically between the singular subspace and
the nonsingular one. This issue might be the source of the mysterious
negative gap that appears in~\cite{rommer-ostlund-1995} in the
vicinity of the critical point.

We have employed a slightly different method for the regularization of
$N_{eff}(k)$. Instead of projecting the problem into the strictly
non-singular subspace, we restrict ourselves to the subspace
in which the eigenvalues of $N_{eff}(k)$ are larger than some $\epsilon$.
There is a tradeoff between loss in precision due to this projection
and loss in precision due to the bad conditioned generalized eigenvalue
problem. In the end we have settled for a seemingly optimal
$\epsilon=10^{-11}$.

\section{Numerical results}
\label{sec:results}

We have applied the algorithm presented above to two exactly solvable
nearest neighbour interaction spin models in order to benchmark its
accuracy: the quantum Ising model and the Heisenberg spin-$1/2$
antiferromagnetic chain.
Even though the Heisenberg model is exactly solvable by
means of Bethe ansatz, in practice it is much harder to obtain its
entire low-energy spectrum.
This is due to the fact that the elementary excitations
are two-spinon states and among these, the solution of the Bethe ansatz
equations for the two-spinon singlet states are computationally
very challenging~\cite{INTRO2BA-II}. Thus for long chains we have restricted
ourselves to check only the precision of the lowest two-spinon triplets.
For small chains that are accessible via exact diagonalization on the
other hand, we compare not only the entire low-energy spectrum
but also the fidelity of the states themselves.


\begin{figure*}[ht]
  \begin{center}
    \includegraphics[width=1.0\textwidth]{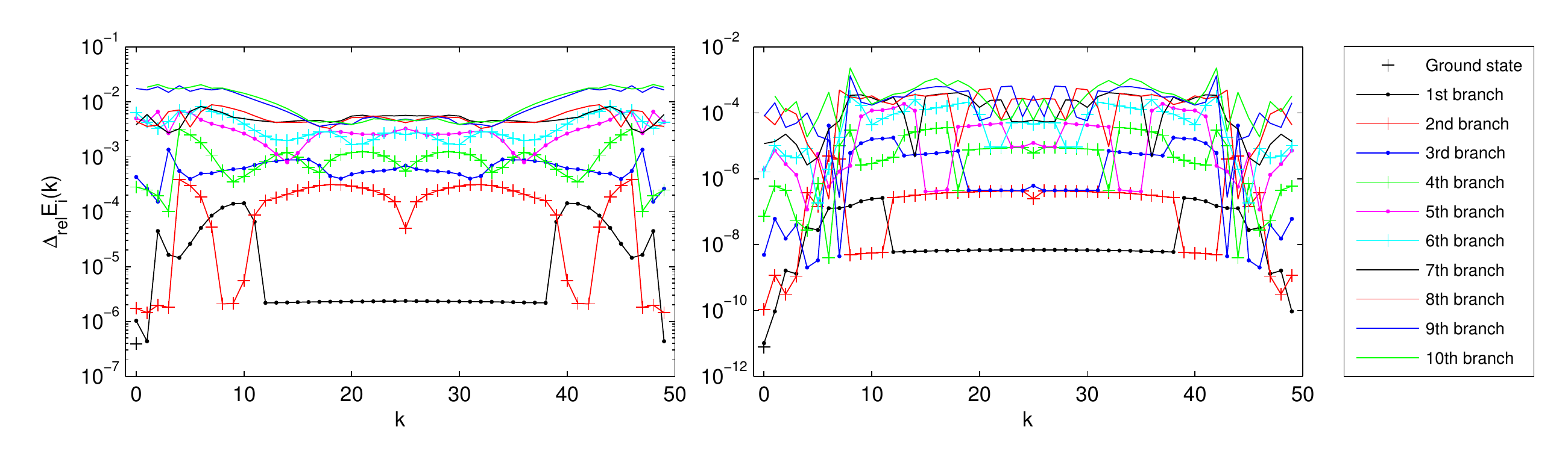}
  \end{center}
  \caption{
    (Color online).
    Relative precision of the low excitation spectrum for a critical
    Ising chain with $N=50$. Left: $D=8$.
    Right: $D=32$.
  }
\label{fig:IS_reldiffEi_g10_N50}
\end{figure*}

\begin{table*}[ht]
  \begin{center}
    \includegraphics[width=1.0\textwidth]{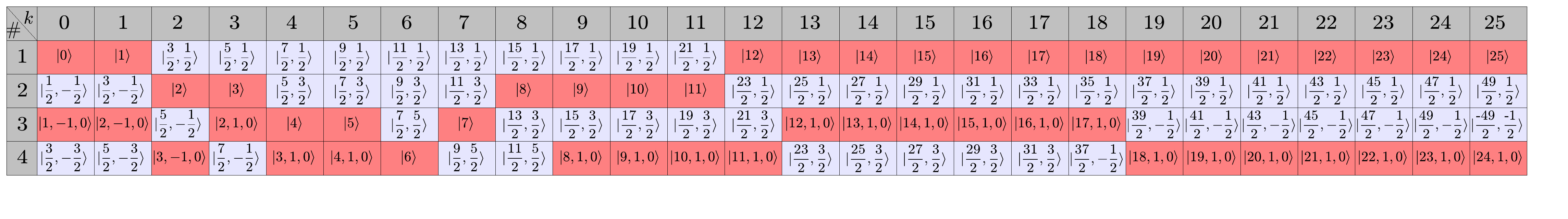}
  \end{center}
  \caption{
    (Color online).
    Quasi-particle structure of the lowest four branches for $g=1$.
    The red/blue (grayscale: dark/light) boxes highlight states from the
    odd-parity subspace respectively from the even parity subspace.
    The ground state which is not shown in the table is the
    fermionic vacuum in the even parity subspace
    i.e. $\ket{GS}=\ket{\Omega}_{even}$.
  }
\label{tab:IS_g10_N50_spectrum_from_elem_exc}
\end{table*}

\begin{figure}[ht]
  \begin{center}
    \includegraphics[width=1.0\columnwidth]{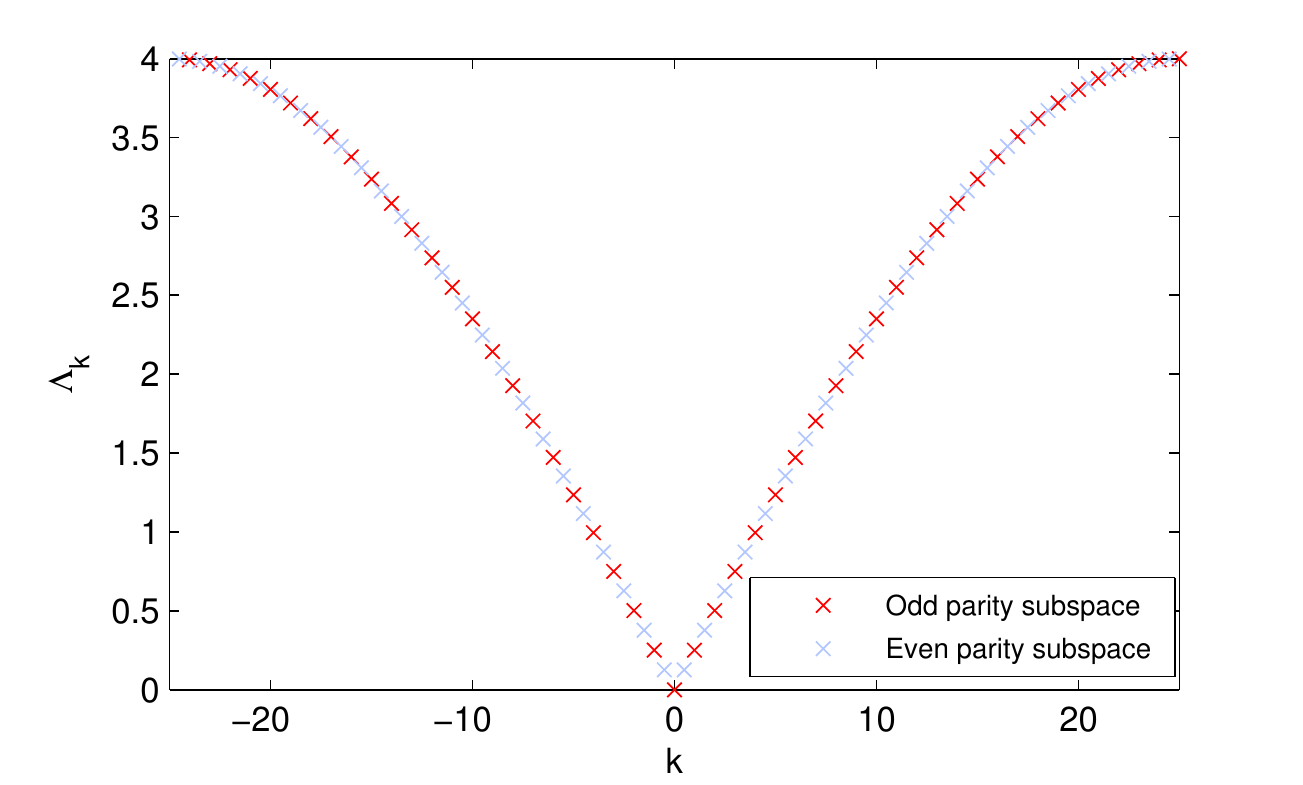}
  \end{center}
  \caption{
    (Color online).
    Exact solution for the dispersion relation of the Bogoliubov modes at
    criticality (i.e., $g=1$).
  }
\label{fig:IS_LAMBDAk_g10_N50_EXACT}
\end{figure}

\subsection{Quantum Ising model}
The Hamiltonian we have used in our simulations of the Quantum Ising
model is given by

\begin{equation}\label{eq:H_IS_sim}
  H_{IS} = -\sum_{i} Z_i Z_{i+1} - g\sum_{i}X_i
  \,\,\,.
\end{equation}
\noindent
We have used this version rather than

\begin{equation}\label{eq:H_IS_exact}
  H'_{IS} = -\sum_{i} X_i X_{i+1} - g\sum_{i}Z_i
  \,\,\,.
\end{equation}
\noindent
due to the fact that having a diagonal interaction term is more convenient
for the numerics.
Of course both versions are equivalent since they can be transformed into
each other by means of the unitary transformation
$U=\bigotimes_{i=1}^{N}H_i$ where the $H_i$ are 1-qubit Hadamard gates.
For the exact diagonalization however we have used (\ref{eq:H_IS_exact})
in order to stick closer to the procedure given in
\cite{lieb-schultz-mattis-1961} where the authors treat the
spin-$1/2$ XY chain.

The diagonalization of (\ref{eq:H_IS_exact}) for PBC in the limit of an
infinite number of sites is straightforward~\cite{pfeuty-1970}.
The first thing one has to do is to map the spin Hamiltonian to a fermionic
one via a Jordan-Wigner transformation.
Now the Jordan-Wigner transformation is non-local due to the fact
that it transforms local spin operators into fermionic ones
that anticommute if they act on different sites.
Luckily for almost all terms in the Hamiltonian the non-localities
cancel except for the term representing the interaction between
the last and the first site. This term ends up containing a global
parity operator acting on all sites and thus breaking the translational
invariance with respect to the fermionic modes.
Now, if we are interested in the thermodynamic limit, we will eventually
take the limes $N\to\infty$ at some point, and in this limit the
contribution of one interaction term to the energy can be neglected.
We thus have the freedom
to alter this term as we please in order to simplify things. One very
convenient choice are the so-called Jordan-Wigner boundary conditions
which are nothing more than simply neglecting the global parity operator
in the last term thereby yielding the fermionic Hamiltonian translationally
invariant. Note that the Jordan-Wigner boundary conditions cannot be
expressed in a trivial way in terms of spin operators.

The fermionic Hamiltonian obtained in this way is quadratic and
translationally invariant, but it is not particle conserving.
This can be fixed by a canonical transformation
\cite{lieb-schultz-mattis-1961} to non-interacting
Bogoliubov fermions. The ground state of the system is then given by
the new fermionic vacuum while excited states can be obtained by
sequentially filling the fermionic modes.
Ordering the eigenstates of the original spin model by momentum
and energy, it turns out that the lowest energy branch coincides
with the dispersion relation of the Bogoliubov fermions.
This happens because for a given
momentum, the lowest energy state is always a state where precisely
one fermionic mode is occupied.

Now, for finite systems with periodic boundary conditions, the Hamiltonian
after the Jordan-Wigner transformation presents a difficulty:
due to the fact that in this case we cannot choose the boundary conditions
freely, there is one term that contains a global parity operator as
prefactor (see Eq. 2.11' in \cite{lieb-schultz-mattis-1961}).
At first sight, this term makes the Bogoliubov transformation impossible.
However, if we project the Hamiltonian onto the subspaces with either
odd or even parity, we can replace the parity operator by its eigenvalue
in that subspace s.t. it becomes $\pm 1$, and
we can apply the Bogoliubov transformation in each subspace separately.
The spectrum of the original Hamiltonian can then be constructed by
picking from each subspace the states with the correct parity. It turns
out that we can arbitrarily choose the sign of the Bogoliubov
parity by shifting the Fermi surface. For example, if we choose the
fermionic vacuum to be the state with lowest energy, all excited states are
particle excitations
\footnote{
If the fermionic vacuum is not the lowest energy state there also exist
hole and particle-hole excitations. If we choose the vaccum in such a way
that there exists exactly one hole-mode we effecively switch the sign of the parity operator opposed to choosing no hole-modes at all.
}
and it turns out that the parity operator
changes its sign under the Bogoliubov transformation for fields
below the critical point i.e. $g < 1$. For $g \ge 1$ this choice of
the vacuum state leaves the parity operator invariant.
Thus for $g<1$, in principle we can switch the sign of the parity operator
by shifting the Fermi surface and thereby we could always
define the Bogoliubov modes such that the parity operator remains invariant.
We will however give numerical evidence for the fact that choosing the Fermi
surface to be the fermionic vacuum state is the physical choice.

Let us first present the results obtained at the critical point $g=1$.
In Figure~\ref{fig:IS_Ei_g10_N50} we have plotted the energy of the lowest
ten branches of excitations of a chain with $50$ spins obtained for MPS bond
dimensions $D=8$ and $D=32$. The results for $D=32$ are so close to the
exact spectrum that it makes much more sense to look at plots of the
relative energy precision rather than at plots of the energy itself.
This is shown in Figure~\ref{fig:IS_reldiffEi_g10_N50}.

\begin{figure*}[ht]
  \begin{center}
    \includegraphics[width=1.0\textwidth]{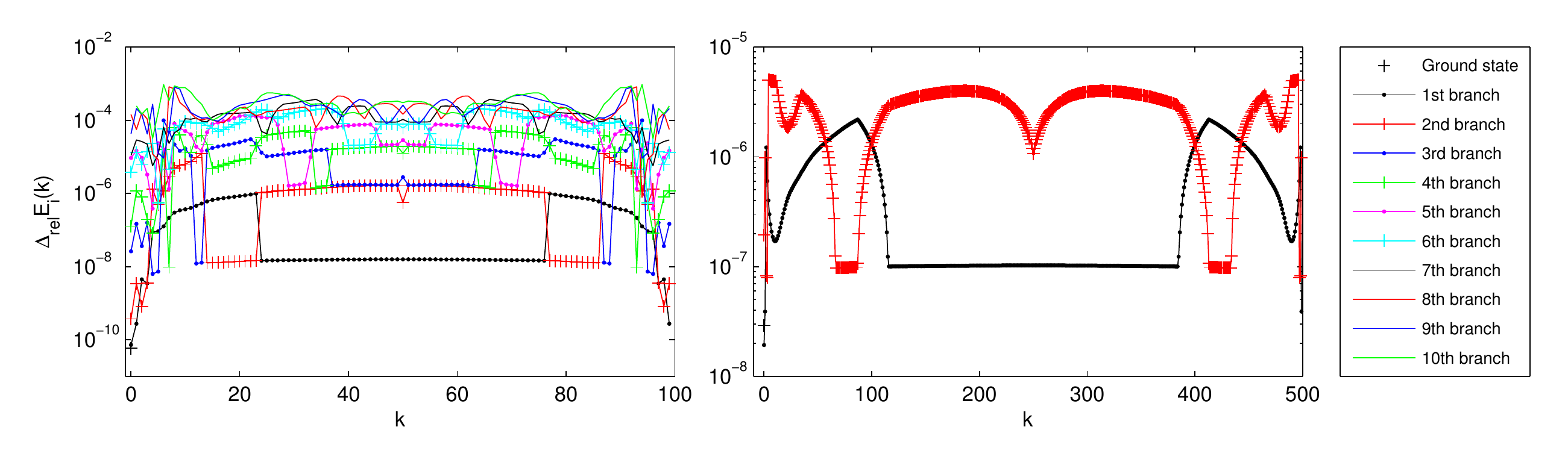}
  \end{center}
  \caption{
    (Color online).
    Relative precision of the low excitation spectrum for the critical
    Ising chain with different chain lengths. Left: $N=100$, $D=32$.
    Right: $N=500$, $D=20$.
  }
\label{fig:IS_reldiffEi_g10_N100_N500}
\end{figure*}

At first sight the crossing of the precision line for the first branch
of excitations with the
one for the second branch seems a little unusual. How can it be that
states with higher energy are approximated by roughly two orders of
magnitude better than states with lower energy? The answer to this question
is obvious if one looks at how the eigenstates emerge from the elementary
Bogoliubov modes. Table~\ref{tab:IS_g10_N50_spectrum_from_elem_exc}
shows which Bogoliubov modes contribute to each individual eigenstate
in the first four branches of excitations. Modes from the even parity
subspace have half-integer momentum while the ones from the odd parity
subspace have integer momentum.
Note that since only excitations with an even number of particles
are allowed in the even-parity subspace, the resulting states always
have integer momentum.
Henceforth $\ket{\Omega}_{even}$ shall denote the vacuum in the even parity
subspace and $\ket{\Omega}_{odd}$ the vacuum in the odd parity one.
The ground state of the critical chain is the Bogoliubov
vacuum in the even parity subspace i.e.
$\ket{GS}=\ket{\Omega}_{even}$.
The first excited state is the zero momentum state from the first branch
and is given by a Bogoliubov mode with zero momentum from the odd
parity subspace
\footnote{
Actually at $g=1$ the zero momentum mode has energy zero so the
states $\ket{\Omega}_{odd}$ and $\ket{0}$ have exactly
the same energy. However this happens only at the critical point $g=1$.
In general $\ket{\Omega}_{odd}$ and $\ket{0}$ have different energy.
}.
It is sufficient to show in
Table~\ref{tab:IS_g10_N50_spectrum_from_elem_exc}
how the spectrum emerges from elementary excitations for momenta
$0\leq k \leq N/2$ since the dispersion relation of the Bogoliubov
fermions is symmetric around $k=0$ as can be seen in
Figure~\ref{fig:IS_LAMBDAk_g10_N50_EXACT}.
The important thing to notice in
Table~\ref{tab:IS_g10_N50_spectrum_from_elem_exc} is that the lowest
branch of excitations does not contain solely one-particle excitations
as it does in the case of the infinite chain. Looking back at the right plot
in Figure~\ref{fig:IS_reldiffEi_g10_N50} we see immediately that
the one-particle excitations from the first two branches are approximated
with roughly the same accuracy between $10^{-11}$ and $10^{-9}$ with the
lower value for small momentum states.
One can easily check that the states with the same order of accuracy from
higher branches are precisely one-particle excitations.
On the other hand it is obvious that two-particle excitations from any
branch where one of the particles has fixed momentum $k=1/2$
can be found in the plateau with relative precision of roughly $10^{-7}$.
The other plateaus of similar precision in the $D=32$ plot of
Figure~\ref{fig:IS_reldiffEi_g10_N50} represent either two-particle
states where each particle has higher momentum or three and more particle
excitations.

\begin{figure*}[ht]
  \begin{center}
    \includegraphics[width=1.0\textwidth]{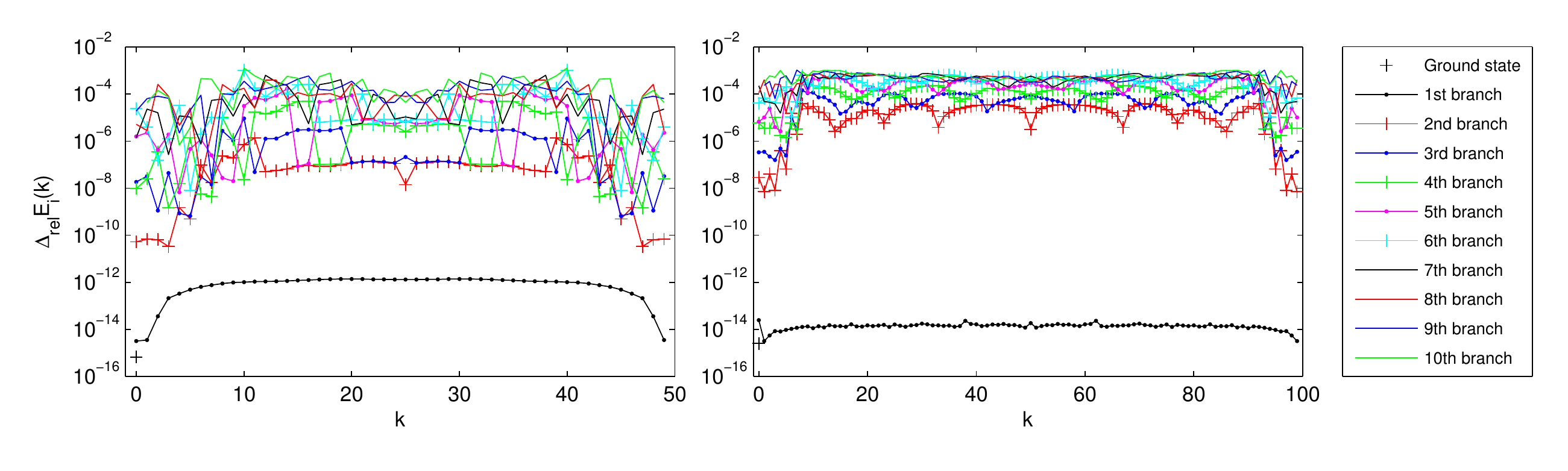}
  \end{center}
  \caption{
    (Color online).
    Relative precision of the low excitation spectrum for the
    Ising chain at $g=1.1$ for different chain lengths. Left: $N=50$, $D=32$.
    Right: $N=100$, $D=32$.
  }
\label{fig:IS_reldiffEi_g11_N50_N100}
\end{figure*}

\begin{table*}[ht]
  \begin{center}
    \includegraphics[width=1.0\textwidth]{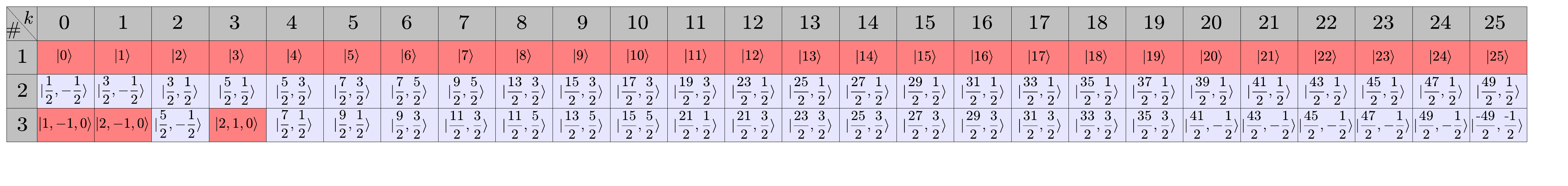}
  \end{center}
  \caption{
    (Color online).
    Quasi-particle structure of the lowest three branches for $g=1.1$.
    The red/blue (grayscale: dark/light) boxes highlight states from the
    odd-parity subspace respectively from the even parity subspace.
    The ground state which is not shown in the table is the
    fermionic vacuum in the even parity subspace
    i.e. $\ket{GS}=\ket{\Omega}_{even}$.
  }
\label{tab:IS_g11_N50_spectrum_from_elem_exc}
\end{table*}

This interpretation of the branch crossings in
Figure~\ref{fig:IS_reldiffEi_g10_N50} is strong evidence for the fact
that (\ref{eq:ansatz}) is a very good ansatz for one-particle excitations.
However it turns out that if $D$ is large enough, (\ref{eq:ansatz}) is
also a fairly good ansatz for many-particle excitations. The reason for
this is that the large bond dimension compensates for the localization
of excitations inherent in ansatz (\ref{eq:ansatz}) by spreading the
effect of the optimized tensor $\B$ to a region around it whose size is
of the order of the induced correlation length of the MPS we start with.
This is exactly why for the Ising chain with $g=1$, $N=50$ and $D=32$
even states from tenth branch are approximated with an accuracy of
roughly $10^{-4}$.

Now let us have a closer look at the region of
the "level-crossing" between the lowest two-fermion branch from
the even parity subspace and the lowest one-fermion branch from the
odd parity subspace. In the case of $g=1$ this crossing turns out to be at
approximately $N/4$. In the immediate neighborhood of the crossing
the energy difference between states with identical momentum becomes very
small. Now if the bond dimension $D$ is chosen such that the precision
of the MPS is of the same order like the interlevel spacing, these two
levels cannot be discriminated properly by the MPS algorithm and thus
there is no clear interpretation we can give to these MPS states in terms
of one or two-particle states. As can be seen in the $D=8$ plot of
Figure~\ref{fig:IS_reldiffEi_g10_N50}, in this region the first and second
MPS branch interpolate between the one and the two-particle states which we
can safely discriminate. Note that this observation holds only on the
side of the level crossing where the one-particle state has higher energy
than the two-particle state (e.g. at $k\approx N/4$ on the left side of
the crossing). On the other side, the one-particle excitation has lower
energy and our one-particle MPS ansatz is perfectly suited to discriminate
between the first and the second branch even if the precision is smaller
than the actual gap between the levels.

The last thing we would like point out about
Figure~\ref{fig:IS_reldiffEi_g10_N50} is the gap in accuracy between the
states from the second and third branch at momentum $k=N/2$. It turns
out that this is a doubly degenerate state since it can be created
by two different superpositions of elementary excitations with
the same energy namely $\ket{\frac{49}{2},\frac{1}{2}}$ and
$\ket{-\frac{49}{2},-\frac{1}{2}}$. This is the reason why the precision
of the $k=N/2$ state in the second branch is better than that of the
surrounding two-particle states which are not degenerated:
variational algorithms are more precise if they try to approximate
the energy of an entire subspace of the Hilbert space rather than that
of a single state. However since all states generated by our algorithm
are orthogonal, the price we have to pay for the improved precision
in the second branch, is a slightly worse precision of the $k=N/2$
state in the third branch.

With this said, we can present the results we have obtained for different
chain lengths $N$ and different values of the magnetic field $g$.
Figure~\ref{fig:IS_reldiffEi_g10_N100_N500} shows the accuracy of the
algorithm for chains with $100$ and $500$ sites at $g=1$.
The plot for $N=100$ is very similar to the $D=32$ plot from
Figure~\ref{fig:IS_reldiffEi_g10_N50}. At small momenta $6\le k\le 11$
the one-particle excitations lie in the branches $4$ to $6$.
These states are not reliably reproduced by our algorithm within
the precision that is otherwise reached for one-particle excitations.
Presumably this would be fixed by increasing the bond dimension $D$
beyond $32$. However at the moment we cannot go to larger $D$ for $N=100$.
For $N=500$ the maximally accessible bond dimension is $D=20$.
The corresponding plot from Figure~\ref{fig:IS_reldiffEi_g10_N100_N500}
is very similar to the small $D$ plot for $N=50$.
Again in the region of the "level-crossing" between one-particle
and two-particle excitations around $k=N/4$ our algorithm has difficulties
obtaining the maximally reachable precision for the one-particle states.

\begin{figure*}[ht]
  \begin{center}
    \includegraphics[width=1.0\textwidth]{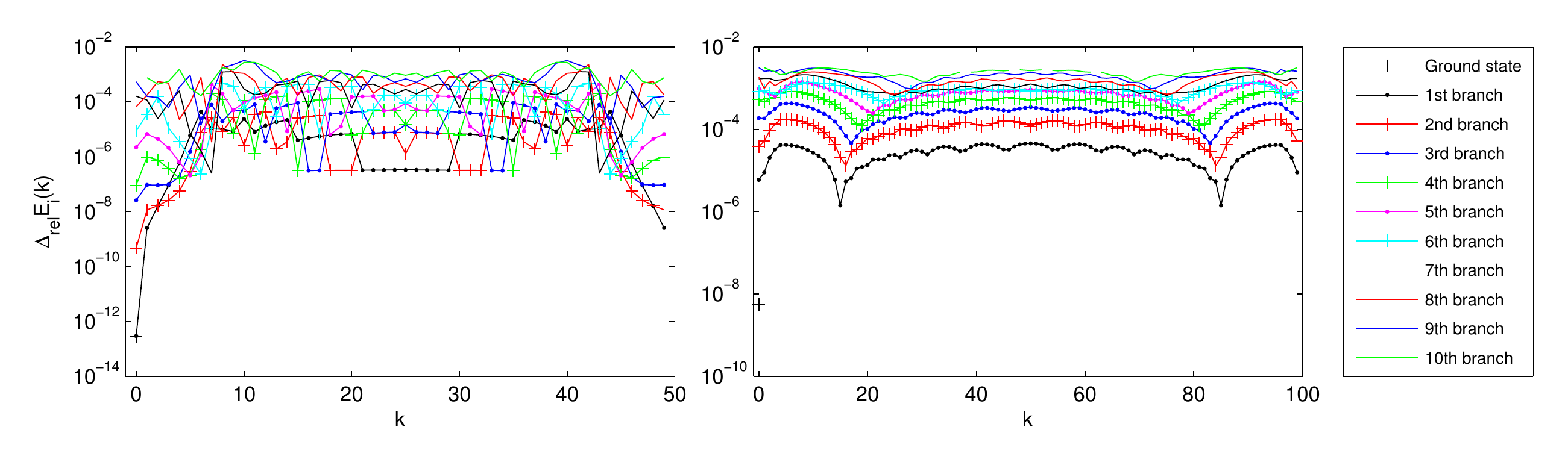}
  \end{center}
  \caption{
    (Color online).
    Relative precision of the low excitation spectrum for the
    Ising chain at $g=0.9$ for different chain lengths. Left: $N=50$, $D=32$.
    Right: $N=100$, $D=32$.
  }
\label{fig:IS_reldiffEi_g9_N50_N100}
\end{figure*}

\begin{table*}[ht]
  \begin{center}
    \includegraphics[width=1.0\textwidth]{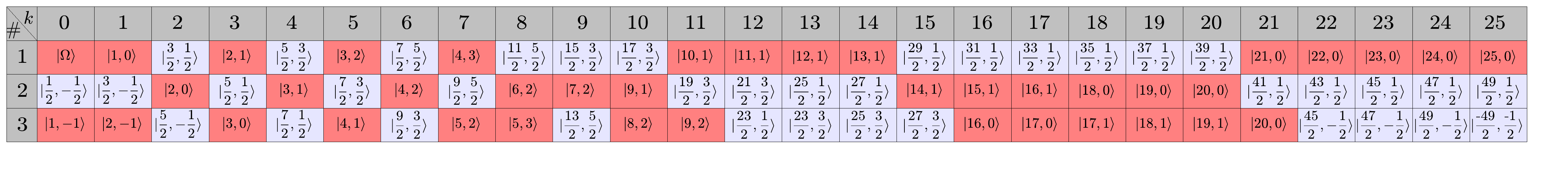}
  \end{center}
  \caption{
    (Color online).
    Quasi-particle structure of the lowest three branches for $g=0.9$.
    The red/blue (grayscale: dark/light) boxes highlight states from the
    odd-parity subspace respectively from the even parity subspace.
    $\ket{\Omega}_{odd}$ denotes the fermionic vacuum in the odd-parity
    subspace. The ground state which is not shown in the table is the
    fermionic vacuum in the even parity subspace
    i.e. $\ket{GS}=\ket{\Omega}_{even}$.
  }
\label{tab:IS_g9_N50_spectrum_from_elem_exc}
\end{table*}

Now let us look at how the algorithm performs when we move away from
the critical point.
Figure~\ref{fig:IS_reldiffEi_g11_N50_N100} shows the relative energy
difference of the MPS approximation for $g=1.1$ i.e. above the critical
point. The most striking feature in this regime is clear separation
of the lowest branch of excitations from the higher ones.
This happens due to the fact that in this case the lowest branch
contains only one-particle states as can be seen in
Table~\ref{tab:IS_g11_N50_spectrum_from_elem_exc}.
Again if $D$ is large enough (e.g. $D=32$ for $N=50$), the different
plateaus of similar precision become clearly visible. The first one
at roughly $\Delta_{rel}E_i(k)\approx 10^{-8}$ contains two-particle
excitations from the second and third branch where one of the
fermionic modes has momentum $k=1/2$. The second one with precision
around $10^{-6}$ consists of states where one of the fermionic
modes has momentum $k=3/2$. Note that in the plot for $N=100$ the
lowest branch has slightly better precision than the one in the
$N=50$ plot even though the virtual bond dimension is the same.
This happens presumably because in this case the chain is long enough
such that the running particle cannot "feel its own tail" due to the PBC.
This is another piece of evidence that ansatz (\ref{eq:ansatz})
is very well suited to describe one-particle excitations. Whether
many-particle excitations are faithfully reproduced depends strongly
on the magnitude of $D$ with respect to $N$.

For $g<1$ the picture changes dramatically.
We can see in figure~\ref{fig:IS_reldiffEi_g9_N50_N100} that at $g=0.9$
the best precision for states from the lowest branch is five to seven
orders of magnitude worse than for $g=1.1$. Without any knowledge of the
quasi-particle structure of the spectrum this huge difference might look
a bit surprising. Even more surprising is the fact that the best precision
at $g=0.9$ is one order of magnitude worse than at the critical point $g=1$.
However looking at the quasi-particle structure in
Table~\ref{tab:IS_g9_N50_spectrum_from_elem_exc} clarifies the
situation. As already mentioned above the parity of the Bogoliubov
fermions in the odd-parity subspace can in principle be arbitrarily
chosen by shifting the Fermi surface. Throughout this work have made the
most natural choice of choosing all modes to have positive energy i.e.
none of the quasi-particle excitations are hole modes. For $g<1$ this choice
switches the sign of Bogoliubov parity operator such that we must
pick states with an even number of excitations from the odd-parity subspace.
One might argue against this convention and claim that it would be
much more natural to pick the Fermi surface s.t. the zero-momentum
mode is a hole excitation which yields the Bogoliubov parity operator
identical to the spin parity operator. In this case we would
have to construct all states from this subspace using an odd number
of quasi-particles. On the other hand
Table~\ref{tab:IS_g9_N50_spectrum_from_elem_exc}
clearly shows that our one-particle excitation ansatz (\ref{eq:ansatz})
is a poor approximation to all states in this regime thereby indicating
that indeed for $g<1$ there exist no one-particle excitations.
Thus our choice of the Fermi surface is justified and we have
to construct the spectrum by picking the even quasi-particle excitations
from the odd-parity subspace.

We can understand this behavior from another point of view if we consider
an infinite chain with open boundary conditions. It is well known that
in the region of the phase diagram where the ground state is
doubly degenerated, the elementary excitations are kink excitations.
If we would however impose
periodic boundary conditions on the infinite chain, the single kink states
would not be eigenstates any more since the existence of one domain
wall would automatically imply the existence of a second one.
In finite systems with PBC, the situation is a bit more complicated
since the ground state degeneracy is not exact
(the energy difference decays exponentially with $N$),
but we can still argue along similar lines that localized
perturbations, that interpolate between the states of the
almost degenerated ground state manifold, must always come in pairs.

\subsection{Heisenberg model}

The other model we have studied is the antiferromagnetic (AF)
Heisenberg spin-$1/2$ chain.
The Hamiltonian reads

\begin{equation}\label{eq:H_HB}
  H_{HB} = \sum_{i=1}^{N}  \vec{S}_i \vec{S}_{i+1}
  = \frac{1}{4} \sum_{i=1}^{N}
   (\sigma^x_i \sigma^x_{i+1} + \sigma^y_i \sigma^y_{i+1}
   +\sigma^z_i \sigma^z_{i+1})
\end{equation}
\noindent
where $S^\alpha=\sigma^\alpha/2$ and $\sigma^\alpha$ denote as usually
the Pauli operators. As we already mentioned, the tensors
$\A$ that constitute the backbone of ansatz (\ref{eq:ansatz}) are
the results of the simulations presented in~\cite{me-2010-PBCI}.
In that work we have obtained a TI MPS approximation of ground states
for finite spin chains with PBC using matrices $A_i$ that were
real and symmetric. These results themselves were based
on previous work~\cite{me-2010-MPOR} where we have approximated the
ground state of infinite OBC chains by TI MPS with real symmetric matrices.
Thus the starting point in the entire procedure that leads ultimately
to the excited states presented here is the imaginary time evolution
for an infinite chain
with a set of real symmetric matrix product operators (MPO).
As we explained in~\cite{me-2010-MPOR} it is not possible to
construct these directly from the the Hamiltonian (\ref{eq:H_HB}).
However, by means of the unitary transformation
$U=U^{\dagger}=\prod_{j=1}^{N/2} \sigma^y_{2j-1}$
(i.e. acting with a $\sigma^y$-gate on every second site) we obtain

\begin{equation}\label{eq:H_HB_transf}
  H'_{HB}=U^{\dagger} H_{HB} U
  =\frac{1}{4} \sum_{i=1}^{N}
   (-\sigma^x_i \sigma^x_{i+1} + \sigma^y_i \sigma^y_{i+1}
    -\sigma^z_i \sigma^z_{i+1})
\end{equation}
\noindent
which allows us to express the imaginary time evolution in terms
of real symmetric MPO. Note that in order for this procedure to work
we have to restrict ourselves to chains with an even number of sites.
In this case it does not matter if we apply the $\sigma^y$-gates
on sites with an odd or an even index, so without loss of generality
we will apply them on the odd ones.
Now $H_{HB}$ and $H'_{HB}$ have the same spectrum and since their
eigenstates are simply related to eachother by
\begin{equation}\label{eq:eigenstate_transf}
  \ket{\psi_i}=\prod_{j=1}^{N/2} \sigma^y_{2j-1} \ket{\psi_i'}
\end{equation}
\noindent
we can digonalize $H'_{HB}$ first and obtain the eigenstates of $H_{HB}$
subsequently with very little effort.

\begin{table}[ht]
  \begin{center}
    \includegraphics[width=0.8\columnwidth]{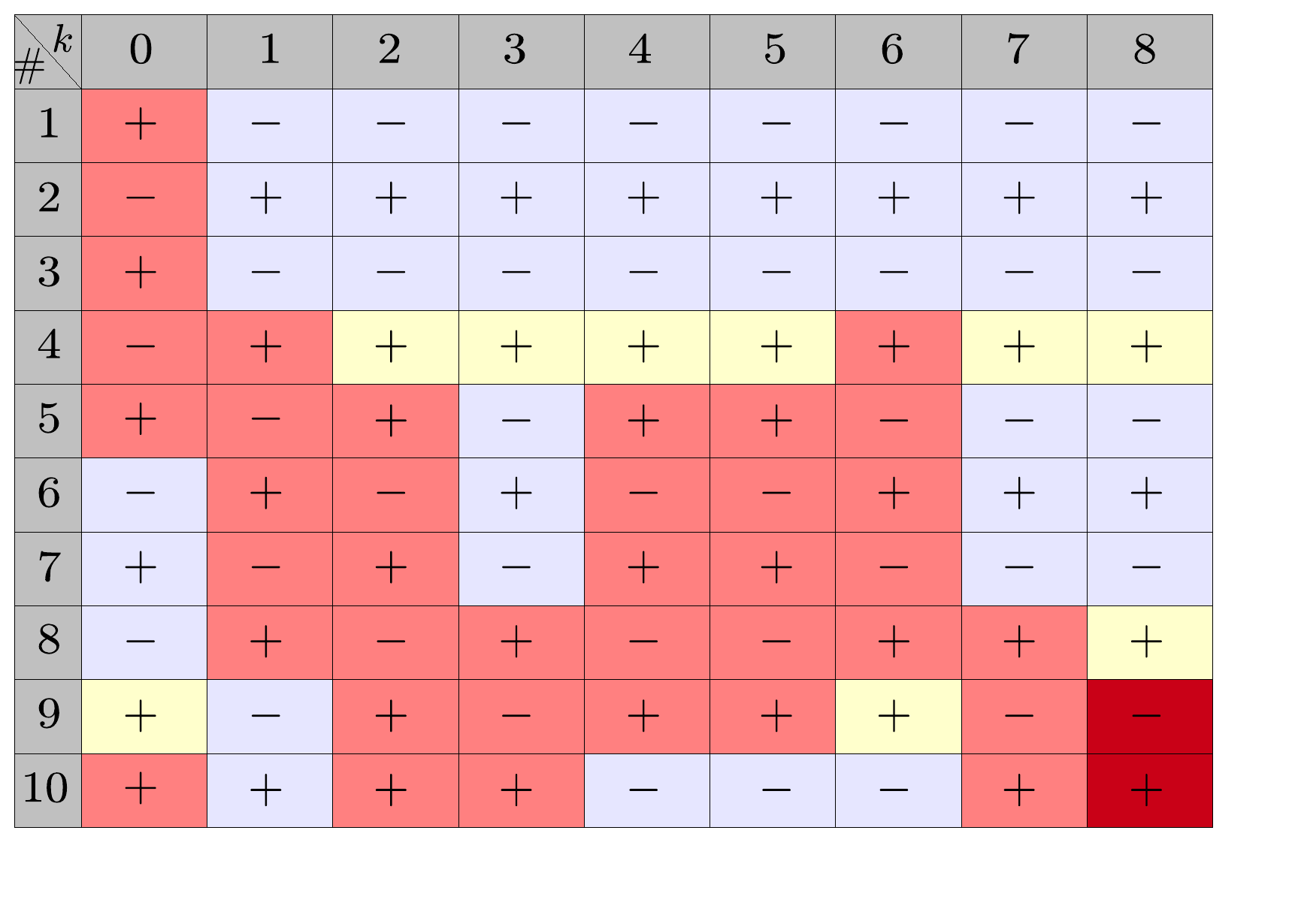}
  \end{center}
  \caption{
    (Color online).
    Multiplet structure of the lowest ten branches of excitations
    for a Heisenberg
    $16$-site chain with Hamiltonian (\ref{eq:H_HB}).
    The colors encode the multiplet information: yellow-singlet,
    blue-triplet, red-quintuplet, dark red-septuplet
    (grayscale: darker colors encode higher multiplets).
    The states within each multiplet are ordered according to their
    total spin projection quantum number.
    The sign denotes the parity of a state.
  }
\label{tab:HB_N16_spectrum_multiplets_parity}
\end{table}

\begin{table}[ht]
  \begin{center}
    \includegraphics[width=0.8\columnwidth]{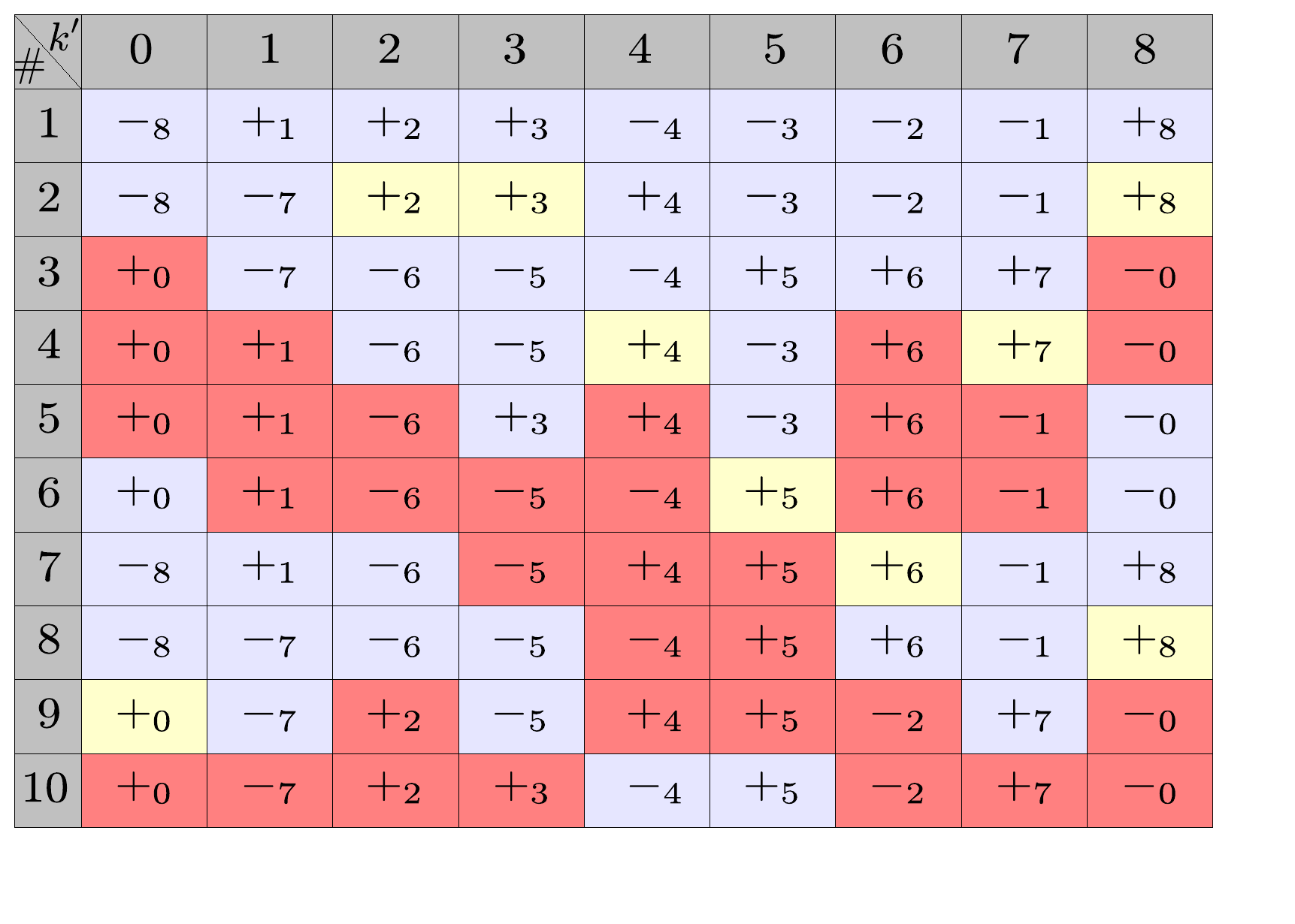}
  \end{center}
  \caption{
    (Color online).
    Multiplet structure of the lowest ten branches of excitations
    for a Heisenberg
    $16$-site chain with Hamiltonian (\ref{eq:H_HB_transf}).
    The colors encode the multiplet information: yellow-singlet,
    blue-triplet, red-quintuplet
    (grayscale: darker colors encode higher multiplets).
    The sign denotes the parity of a state and the index
    denotes the momentum $k$ if we apply the transformation
    (\ref{eq:eigenstate_transf}) to an eigenstate with momentum  $k'$.
  }
\label{tab:HBtransf_N16_spectrum_multiplets_parity}
\end{table}

We will show below that the momentum of a state is not always invariant under
the transformation (\ref{eq:H_HB_transf}).
The easiest way to obtain the momentum for any given state is by
computing the expectation value of the translation operator $T$ with
respect to that state.
$H_{HB}$ and $H'_{HB}$ are both translationally invariant thus
all their eigenstates have well defined momentum so we can be
sure that the reverse transformation
$\ket{\psi_i'(k')} \rightarrow \ket{\psi_i(k)}$ will map momentum
eigenstates to momentum eigenstates albeit $k$ will generally differ
from $k'$.
The relation between the momenta follows easily from

\begin{equation}\label{eq:k_transf}
  \begin{split}
    &e^{-i\frac{2\pi k}{N}}=\braket{\psi_i(k)|T|\psi_i(k)} =\\
    &=\braket{\psi_i'(k')|\bigg(\prod_{j=1}^{N/2} \sigma^y_{2j-1}\bigg)
              T \bigg(\prod_{j=1}^{N/2} \sigma^y_{2j-1}\bigg) |\psi_i'(k')} \\
    &=\braket{\psi_i'(k')|\bigg(\prod_{j=1}^{N/2} \sigma^y_{2j-1}\bigg)
              \bigg(\prod_{j=1}^{N/2} \sigma^y_{2j}\bigg) T |\psi_i'(k')} \\
    &=e^{-i\frac{2\pi k'}{N}}
      \braket{\psi_i'(k')|\prod_{j=1}^{N} \sigma^y_{j}\,|\psi_i'(k')}
     =e^{-i\frac{2\pi k'}{N}} \braket{P_y}_{i',k'}
  \end{split}
\end{equation}
\noindent
where we have used $T \big(\prod_{j} O_{j}\big) T^{-1}=\prod_{j} O_{j+1}$
and $T\ket{\psi_i'(k')}=e^{-i\frac{2\pi k'}{N}}\ket{\psi_i'(k')}$.
Thus the change in momentum depends solely on the expectation value of
the operator $P_y=\prod_{j=1}^{N} \sigma^y_{j}$ which in the following
we will call the \emph{parity} operator. This naming convention
makes sense since $P_y=i^N\exp(i \pi S_T^y)$ where
$S_T^y=\sum_{j=1}^N S_j^y$ thus $P_y$ measures the parity of the
total spin along the $y$-direction. Note that due to the factor $i^N$
the meaning of positive and negative parity is interchanged for chains
with $N=0(\mod{4})$ and chains with $N=2(\mod{4})$.
The parity is a good quantum number for both $H_{HB}$ and $H_{HB}'$
so there exist eigenstates $\ket{\psi_i'(k')}$ that have well defined
parity plus or minus one. If $\braket{P_y}_{i',k'}=+1$ the momentum
remains unchanged i.e. $k=k'$, if
$\braket{P_y}_{i',k'}=-1=e^{\pm i\pi}$ we have
$k=k'\oplus_N N/2$
where $\oplus_N$ denotes addition modulo $N$.
Note that the parity itself is invariant under the mapping between
$H_{HB}$ and $H_{HB}'$ since $U^{\dagger}P_y U=P_y$.

\begin{figure*}[ht]
  \begin{center}
    \includegraphics[width=1.0\textwidth]{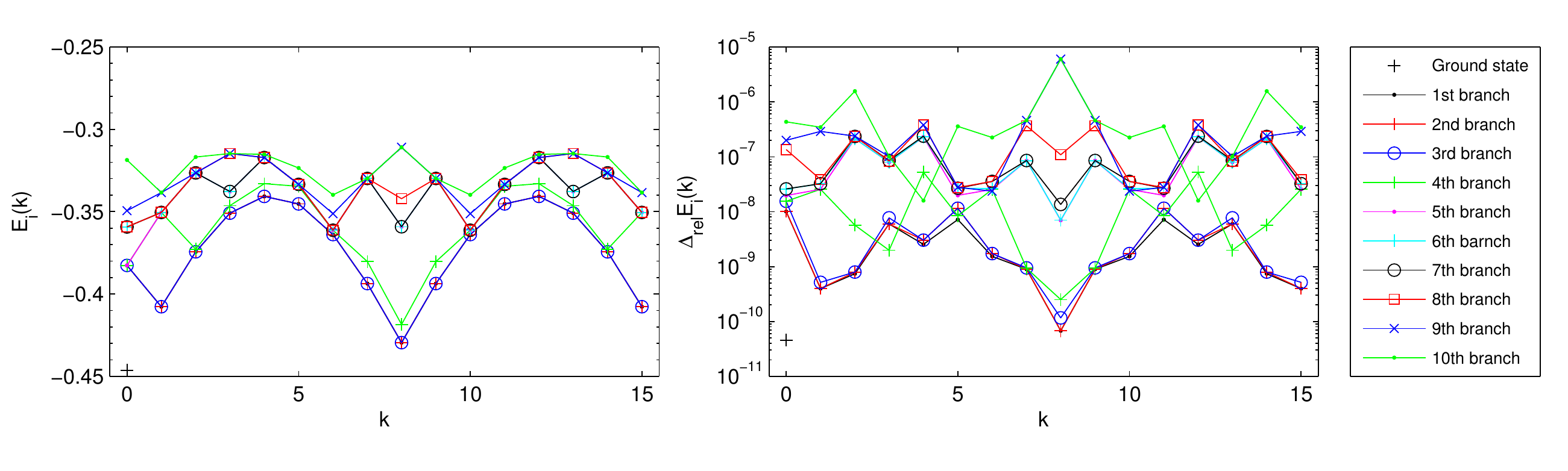}
  \end{center}
  \caption{
    (Color online).
    Results for the low excitation spectrum (left) and the corresponding
    relative precision (right) for the Heisenberg spin-$1/2$ chain with
    $N=16$ sites.
  }
\label{fig:HB_Ei_reldiffEi_N16}
\end{figure*}

\begin{figure}[ht]
  \begin{center}
    \includegraphics[width=1.0\columnwidth]{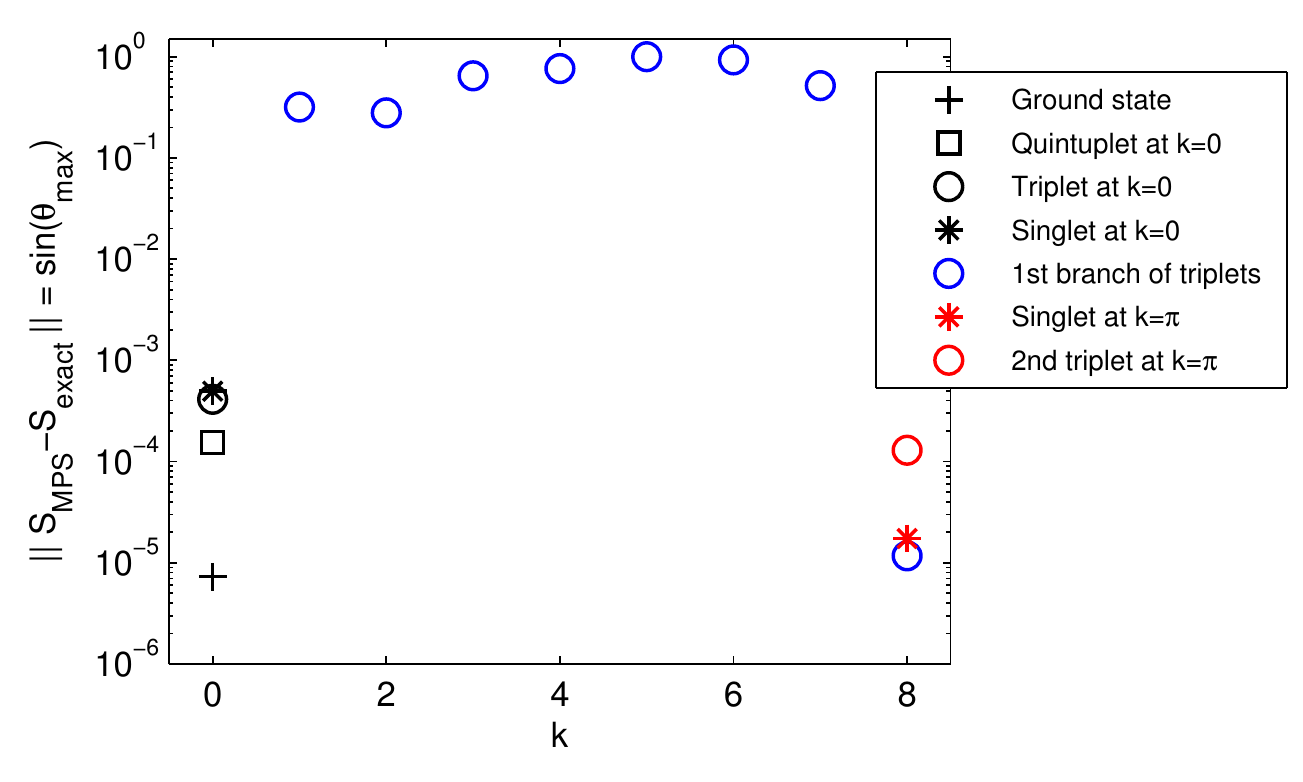}
  \end{center}
  \caption{
    (Color online).
    Distance between several degenerated subspaces obtained by our algorithm
    to the corresponding degenerated subspaces obtained by exact
    diagonalization. As a measure for the distance we have used the sine
    of the canonical angle with the largest magnitude as defined
    in~\cite{book-bhatia-1997}.
  }
\label{fig:HB_N16_canonicalANGLES}
\end{figure}

Now the generators of the $SU(2)$ symmetry for $H_{HB}'$ do not
commute with the translation operator thus we cannot classify
the momentum eigenstates in terms of irreducible representations
of $SU(2)$. For $H_{HB}$ however we can do this, so we
know exactly the degeneracy structure of the spectrum in each
subspace with fixed momentum.
Thus if we encounter for instance a threefold degenerated eigenstate of
$H_{HB}'$, we know this is mapped to a spin triplet with
well defined momentum in the original Hamiltonian.
Accordingly it must contain a two-dimensional subspace with
negative parity corresponding to total spin along the $y$-direction
$\pm 1$ and a one-dimensional subspace corresponding to total spin $0$.
Since the spin triplet in the original Hamiltonian has well defined
momentum, according to the rules for the mapping $k\leftrightarrow k'$, we
will have one eigenstate of $H_{HB}'$ with momentum $k$ and a
two-dimensional subspace with the same energy but different momentum
$k'=k\ominus_N N/2$.
In this way
\footnote{
A singlet state would have parity $+1$ and thus there would be no
momentum shift in this case. A quintuplet would contain a three-dimensional
subspace with parity $+1$ and a two-dimensional subspace with parity
$-1$. Thus in this case we would observe three states with no momentum
shift and two states with a $\pi$-shift. The generalization
to higher multiplets is obvious.
},
after approximating the spectrum of $H_{HB}'$ and labeling
all energies with the corresponding momentum we can obtain
the spectrum of $H_{HB}$ by mere inspection of the degeneracy structure.
Table~\ref{tab:HB_N16_spectrum_multiplets_parity} and
Table~\ref{tab:HBtransf_N16_spectrum_multiplets_parity} illustrate
how the multiplets of $H_{HB}$ and $H_{HB}'$ are related to eachother.

\begin{figure*}[ht]
  \begin{center}
    \includegraphics[width=1.0\textwidth]{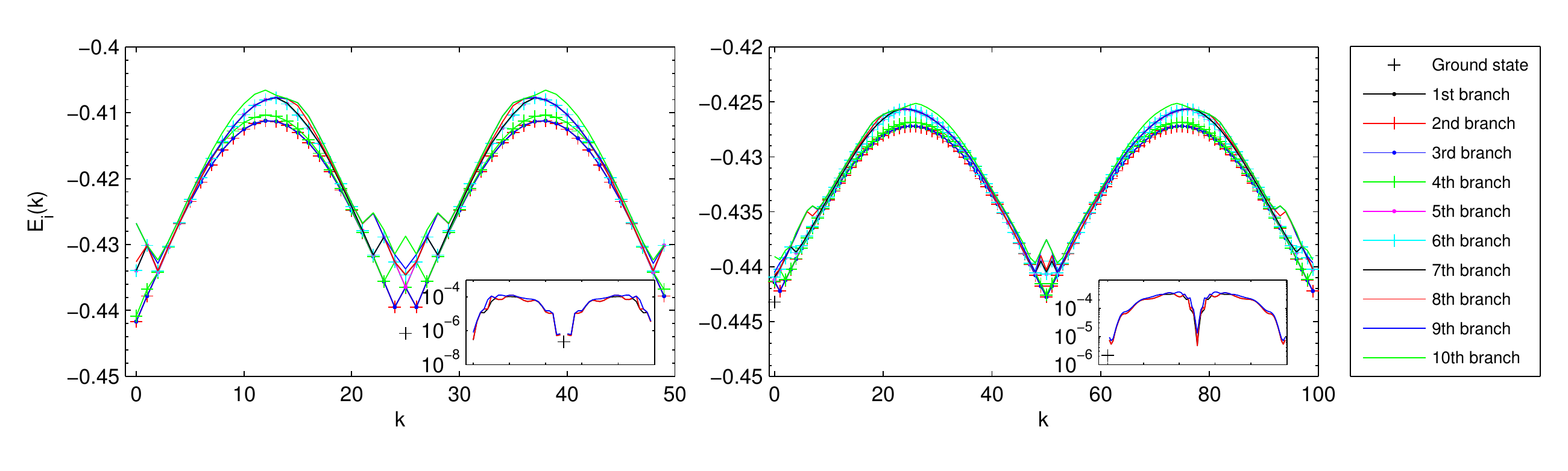}
  \end{center}
  \caption{
    (Color online).
    Results for the low excitation spectrum and the corresponding
    relative precision of the lowest triplet (insets) for the Heisenberg
    spin-$1/2$ chain with $N=50$ (left) and $N=100$ (right) sites.
  }
\label{fig:HB_Ei_reldiffEi_N50_N100_D32}
\end{figure*}

This procedure works very well for the lower branches of the
dispersion relation where the precision of our simulation is good
enough to discriminate unambigously between different multiplets.
For higher branches, on one hand the precision gets worse and
on the other hand the density of states increases such that
multiplets with similar energy become effectively undistiguishable
for our algorithm. In this case the eigenstates with well defined
momentum that we obtain for the Hamiltonian $H_{HB}'$ do not have
well defined parity i.e. they mix parity eigenstates with different
parity. Since states with same momentum and different parity are mapped by
(\ref{eq:eigenstate_transf}) to states with different momentum,
if we start with such a momentum eigenstate we obtain after the
transformation a superposition of states with different momenta which
is clearly not a momentum eigenstate.
There are however two ways to overcome this issue and obtain approximations
of the eigenstates of $H_{HB}$ that are at the same time exact momentum
eigenstates.

The \emph{first} one amounts to computing the matrix elements
of the translation operator $T$ in the subspace spanned by the transformed
states $\{M_{odd}^{y} \ket{\psi_i'(k)'}\}$ where
$M_{odd}^{y}:=\prod_{j=1}^{N/2} \sigma_{2j-1}^{y}$ and
then diagonalize this matrix.
It is not difficult to check that this can be done for each momentum $k'$
separately since
$M_{odd}^{y}\,T\,M_{odd}^{y}=M_{odd}^{y}\,M_{even}^{y}\,T=P_y T$ which
is a translationally invariant operator and thus it does not mix
states with different momentum.
Diagonalizing each of the
$T(k')_{ij}=U^{\dagger}(k')_{il}D(k')_{lm} U(k')_{mj}$
yields for each $k'$ a unitary $U(k')$ that is nothing more than
the transformation that we need to obtain the desired momentum eigenstates
via $\ket{\psi_i(k_i)}= U(k')_{ij} \ket{\psi_j'(k')}$.
The new momentum $k_i$ can be read off the diagonal matrix $D(k')$.
There are two drawbacks that come with this procedure. The first one is
that we must compute the matrix elements $T(k')_{ij}$ each of which is
done with the computational cost $O(N D^5)$. Since there are
$N b^2$ of these where $b$ is the number of branches, we obtain
the overall cost $O(N^2 b^2 D^5)$. Usually we compute enough
branches such that $b^2>D$ holds, thus the cost for this procedure
ends up being higher than the one for the diagonalization of $H_{HB}'$.
The second drawback is that the superpositions
$U(k')_{ij} \ket{\psi_j'(k')}$ mix the original approximations
of the energy levels thereby
slightly lowering the energy of higher excitations but increasing the
energy of lower excitations, which are usually the ones we are most
interested in.

The \emph{second} way to approximate the eigenstates of the
original Hamiltonian $H_{HB}$ such that they are at the same time
exact momentum eigenstates is to add to $H_{HB}'$ a perturbation
that splits degenerated levels with different parity.
This is easily achieved by taking $H_{HB}^{\pm}:=H_{HB}'\pm \lambda P_y$
where $\lambda$ must be chosen such that it is big enough for our
algorithm to deliver only states with a single parity, but as small
as possible in order to avoid numerical inaccuracies caused by
altering the Hamiltonian. In the case of the Heisenberg model, if we
choose to compute $b=10$ branches, $\lambda=0.1\cdot N$ fulfills these
requirements. In practice we first apply our algorithm to
$H_{HB}^{-}$ which yields for each momentum $k'$ $b$ states
with positive parity. These states do not change their momentum
under the transformation (\ref{eq:eigenstate_transf}).
Subsequently we apply the algorithm to $H_{HB}^{+}$ which yields
states with negative parity that change their momentum after
the transformation according to $k=k'\oplus_N N/2$.
In this way we end up with $2b$ branches of states that approximate
the spectrum of $H_{HB}$ and that are at the same time exact momentum
eigenstates. The computational cost per state is thus exactly the
same like diagonalizing only $H_{HB}'$.

Let us first look at the results we have obtained for a small chain with
$16$ sites. We have chosen to look at such a small system first for two
reasons: \emph{First}, even though the Heisenberg model is exactly
solvable via Bethe ansatz, obtaining \emph{all} energy levels can
be quite involved. Choosing $N$ as small as $16$ allows us to
compute the spectrum of this small chain by means of exact diagonalization.
\emph{Second}, even for the energy levels that are easily computable
with the Bethe ansatz (i.e., the triplet states in the subspace of
two-spinon excitations~\cite{INTRO2BA-II}) it is not possible to obtain
the eigenstates themselves. Exact diagonalization of a small chain
on the other hand allows us to compute and store the exact eigenstates
in order to check the fidelity of our MPS approximation.

Figure~\ref{fig:HB_Ei_reldiffEi_N16} shows the energy of the first ten
branches of excitations and the corresponding relative precision.
Note how states belonging to the same multiplet have very similar precision
even though they have different parity and thus correspond to eigenstates
of $H_{HB}'$ with different momentum. Since there are no one-particle
excitations in the low-energy spectrum of the AF Heisenberg model,
we do not obtain such a good precision like in the case of the
quantum Ising model. Nevertheless we get a very good approximation
of the first excited level, namely the triplet excitation at $k=N/2$.
We have also tested the accuracy of the states themselves:
for non-degenerated states, the absolute value of the overlap is a
perfect measure for this, and for reasons that will become clear
immediately, we have looked at the sine of the fidelity.
For degenerated states, in order to compare the
subspace spanned by our MPS to the one spanned by the exact eigenstates,
we have used as a measure for the distance the definition given in chapter
7 of~\cite{book-bhatia-1997}: the sine of the largest canonical angle
between the two subspaces. The canonical angles can be easily computed
from the matrix that has as its entries the overlaps between all states
of the subspaces that we want to compare. The results are plotted in
figure~\ref{fig:HB_N16_canonicalANGLES}. We see that only the MPS
with momentum $k=0$ and $k=N/2$ are extremely accurate. All other
states, especially those around $k=N/4$, are much further away from
the exact solutions, which is a bit surprising given the fact that
the energy precision for these states is comparable to the one obtained
for $k=0$.

The spectrum that we obtain for longer chains is plotted in
figure~\ref{fig:HB_Ei_reldiffEi_N50_N100_D32}. In this
regime we have only looked at the precision of the lowest two-spinon
triplet for which the exact results were obtained following
\cite{INTRO2BA-II}. Again we see that the states at momentum
$k=k_0\oplus N/2$ have the best accuracy. We would like to make
two more remarks concerning the chain with $N=50$. First, note
that the ground state has momentum $k_0=N/2$ in this case.
Second, unlike in the case
of $N=100$, where for all momenta $k\neq k_0$ the lowest excitation is
a triplet, we observe that for $N=50$ this does not happen.
Our simulations reveal that at $k\in\{2,3,47,48\}$ the quintuplet excitation
lies below the triplet while at $k\in\{23,27\}$ it is a singlet
that is the lowest lying excitation.


\section{Conclusions and Outlook}
\label{sec:conclusions}
Inspired by previous approaches~\cite{rommer-ostlund-1995,porras-2006}
we have introduced a new method for the simulation of translationally
invariant spin chains with periodic boundary conditions. We have used
an MPS based ansatz that corresponds to a particle-like excitation
with well defined momentum in order to obtain extremely accurate results
for models where the spectrum contains precisely one-particle states.
For states that can be expressed in terms of many quasi-particle
excitations, we still obtain feasible results if the MPS bond dimension
is chosen to be big enough. In the case of the quantum Ising model,
our results indicate that for $g<1$ the spectrum is built up entirely
out of excitations with an even number of quasi-particles.
Generalizations of our approach can go in two directions:
First, it is possible to adjust ansatz~(\ref{eq:ansatz}) in order to
treat infinite systems with open boundary condition, which
we are addressing in~\cite{juthoEXC-2011}.
Second, it seems feasible to generalize our approach to a many-particle
ansatz by using more than one MPS tensor in~(\ref{eq:ansatz}) in order
to define the variational manifold.

\section{Acknowledgements}

We thank G.Vidal, V. Murg, E. Rico and B. Nachtergaele for valuable
discussions. This work was supported by the
FWF doctoral program Complex Quantum Systems (W1210),
the Research Foundation Flanders, the ERC grant QUERG,
and the FWF SFB grants FoQuS and ViCoM.






\end{document}